\documentclass[review]{elsarticle}

\usepackage{hyperref}

\usepackage{color}
\usepackage{ulem}

\journal{Carbon}

\begin{document}

\begin{frontmatter}
\title{
A fundamental mechanism for carbon-film lubricity identified by
means of {\it ab initio} molecular dynamics
}


\author[mymainaddress,mysecondaryaddress]{Seiji Kajita\corref{mycorrespondingauthor}}
\ead{fine-controller@mosk.tytlabs.co.jp}

\author[mysecondaryaddress]{M. C. Righi\corref{mycorrespondingauthor}}
\cortext[mycorrespondingauthor]{Corresponding authors}
\ead{mcrighi@unimore.it}

\address[mymainaddress]{Toyota Central R\&D Labs., Inc., 41-1, Yokomichi,
              Nagakute, Aichi 480-1192, Japan}
\address[mysecondaryaddress]{Istituto Nanoscienze, CNR-Consiglio Nazionale delle
	      Ricerche, I-41125 Modena, Italy}

\begin{abstract}
Different hypotheses
have been proposed to explain the mechanism for the 
extremely low friction
 coefficient of carbon coatings
and its undesired dependence on air humidity.
A decisive atomistic insight is still lacking because
of the difficulties in monitoring what actually happens at the buried sliding interface.
Here we perform large-scale {\it ab initio} molecular dynamics simulations of
both undoped and silicon-doped 
carbon films sliding in the presence of water.
We observe the 
tribologically-induced surface hydroxylation and subsequent formation
of a thin film of water molecules bound to the OH-terminated surface by hydrogen
bonds.
The comparative analysis of silicon-incorporating and clean surfaces,
suggests that this two-step process can be the key phenomenon to provide high slipperiness to the carbon coatings.
The water layer is, in fact, expected to shelter the carbon surface from direct solid-on-solid contact
and make any counter surface slide extremely easily on it.
The present insight into the wettability of carbon-based films 
can be useful for designing new coatings for biomedical and energy-saving applications
with environmental adaptability.
\end{abstract}


\end{frontmatter}


\section{Introduction}

Carbon-based films, as (poly)crystalline
diamond and amorphous diamond like carbon (DLC), 
have attracted great interest from both the industry and
scientific community due to their exceptional physical, chemical, biomedical, mechanical and tribological
properties.\cite{AHLettington, NYang}
Tribologically, carbon films provide some of the lowest known friction and wear coefficients
without any environmental pollution related to their
use.\cite{fontaine,erdemir,grierson,bewilogua}
However, the widespread application of carbon films has been hindered by a long-standing
problem related to the influence of air humidity on their tribological performances.\cite{lancaster,woydt,jpcc-diamond,sirghi,HLi,SGkim,andersson,AErdemir}
In DLC systems, for example,
both highly positive and negative
effects of  humidity 
on the friction coefficients and wear rates
have been reported 
even in the same type of tribological test conditions.\cite{HLi,SGkim}
This uncontrolled behavior is most likely the result of chemical reactions activated
at the buried sliding interface interacting with water molecules.  
The chemical compositions of the carbon film and its hydrogen content 
can deeply alter the surface reactivity and friction.\cite{andersson,AErdemir}
Indeed, friction reduction by moisture has been attributed to surface passivation by water chemisorption,
which reduces the adhesion with a
counter surface;\cite{jpcc-diamond, andersson,konca,qi,zilibotti,
righiprb2007, righilangmuir} 
especially, hydrophilic hydroxyl groups are considered to 
 play a crucial role in reducing the friction. \cite{jpcc-diamond,XWu, KHayashi,DBarros}
In addition to the passivation, recent nano-scale experiments suggested
another possible atomic mechanism 
to explain the 
extremely low friction of carbon-based film,
which is connected to the presence of water molecules
confined at the friction interface.
\cite{sirghi,IBkim,CMatta,spencer-book}
By means of
atomic force microscope (AFM) measurements it has been
uncovered the presence of an adsorbed water layer, few molecules thick,
which may function as boundary lubricant on DLC films.\cite{sirghi} 
A friction force measurement
showed that the viscosity of the confined
water layer is more than $10^8$ times greater than
that of bulk water.\cite{IBkim}
A nanoconfined water layer has been also considered to
explain the lubrication mechanisms of steel by glycerol
\cite{spencer-book,LPottuz,WHabchi} and polymer brushes. \cite{klein}

The formation of a confined water layer seems favored by the
incorporation of  silicon atoms into the carbon matrix in a suitable dosing.\cite{sanchez,donnet,grill}
Si-doped DLC (Si-DLC) and SiC show, in fact, 
more stable and lower friction coefficients in water and humid
environments than undoped DLC.
Initially, the low friction mechanism of the Si incorporating carbon films
was believed to be completely different from that of Si-free
films and attributed
to silica-sol wear debris resulting from the oxidation of silicon fragments by water
vapor.\cite{woydt,oguri,MGkim,Fzao,scharf,chen2001,chen2002}
However, films incorporating a low amount of silicon present very low
friction in spite of the absence of any significant wear.\cite{dohda,varma} Several authors have reported that 
water can dissociate directly at the silicon sites incorporated at the carbon surface
leading to the formation of hydroxyl groups at these sites.\cite{SGkim,mori2,takahashi,kato,XWu2,sen}
By means of first principles calculations of water adsorption
we showed that the hydroxyl termination enhances the surface
hydrophilicity,\cite{seiji-clelia_static}
in agreement with experimental
observations.\cite{yi,borisenko} 
 Kasuya {\it et al.} performed resonance shear measurements for evaluating
the properties of water confined between silica surfaces with
different concentrations of silanol (Si-OH) groups at the surface, 
and observed 
ice-like water on the surface terminated with a high concentration of
Si-OH, which 
could provide lubricity under high normal pressure more than 
a low concentration of Si-OH groups.\cite{kasuya} 
The formation of a structured water layer upon Si-OH termination of the
carbon surface
was also observed by classical molecular dynamics (MD) simulations.\cite{washizu}
 
Despite the growing interest in understanding 
the atomistic mechanisms for the low friction and
wear of carbon-based coatings,
our present understanding is limited
by the difficulty in monitoring the 
buried sliding interface.
A direct access by experiments is, in fact, extremely challenging
and  molecular dynamics simulations based on 
empirical force fields are typically inadequate 
for an accurate description of chemical reactions occurring 
in conditions of enhanced reactivity.
Here we apply large-scale {\it ab initio} MD simulations
that realistically describe the water/surface interaction,\cite{seiji-clelia_static,manelli,galli}
and provide {\it in situ}, real-time monitoring of tribochemistry processes.\cite{zilibotti}
Thanks to a comparative study of Si-incorporating and pure diamond surfaces we highlight
the pivotal role of the stress-assisted surface hydroxylation and the 
subsequent formation of a nano-confined layer of water molecules strongly bound to the surface.

\section{Method}

We perform {\it ab initio} MD simulations based on the Car Parrinello method\cite{cp}
by means of the pseudopotential/plane-waves computer code
included in the QUANTUM ESPRESSO package,\cite{qe} which has been
modified by our group in order to 
simulate tribological systems.\cite{zilibotti}
Interfaces are modeled by periodic supercells  of 
15.1 \r{A} $\times$ 10.1 \r{A} $\times$ 20.0 \r{A}
dimensions, containing two diamond slabs 6 layers thick, with (6$\times$4)
in-plain size. 
The diamond surface presents a (2$\times$1) reconstruction constituted
of dimers that gives rise to alternating rows and trenches of sp$^2$- 
and sp$^3$-bonded carbon atoms.
Bonds with this different hybridizations are 
 also present in DLC,\cite{FMangolini} 
therefore our model aims at mimicking the {\it local} reactive
sites of the DLC surface. The effects of larger-scale features, such as 
steric effects due to the roughness of DLC
surfaces,\cite{KHayashi} are not taken into account in the present model.
The large
number of atoms included in our system (up to 378) and the simulated
time intervals, about 10 ps for each trajectory,
 render our simulations computationally very demanding, in
particular they required about 100 k cpu hours highly parallel supercomputers per each trajectory,
which points out the importance and the complexity of the current work.

The electronic structure is calculated by means of the density functional
theory (DFT) with the Perdew-Burke-Ernzerhof (PBE) approximation for the
exchange correlation functional.\cite{pbe}
The ionic species are described by
ultrasoft pseudopotentials.\cite{vanderbilt}
The electronic wavefunctions are
expanded in a plane-wave basis set with a cut-off energy of 25 Ry, and 
the Brillouin zone is sampled at the Gamma point.

In the CP dynamics, the ionic and electronic degrees of freedom
evolve simultaneously, a fictitious mass of 450 a.u. 
is assigned to the electrons and the equations of motions are
integrated with the Verlet algorithm 
with a time step of 5 a.u.=0.12 fs. 
Deuterium is used instead of hydrogen 
so as to avoid unphysical coupling between electronic and ionic
degrees of freedom in the concurrent execution of the time evolution.
The temperature of the system is kept constant at 300 K 
by means of a Nos\'{e}-Hoover thermostat composed by double chains 
with frequencies of 30 THz and 15 THz.\cite{nose,evans}
An additional Nos\'{e}-Hoover thermostat is
applied on the electronic degrees of freedom 
with frequency 200 THz. Mechanical stresses are applied along the [001]
(load) and [110] (shear) directions.
The water molecules (12 per supercell) are initially positioned 
in the configuration that we
have identified as the most stable between the carbon slabs,
and then the shear stresses and normal pressure  
are gradually applied adopting the same computational scheme as described in Refs.~\cite{zilibotti,compDetails}.

\section{Results}
\subsection{Tribologically-induced hydroxylation of the carbon film}
We consider a sliding interface composed of a Si-incorporating C(001) 
surface partially hydrogenated and a 
fully hydrogenerated H-C(001) countersurface.
Si atoms are incorporated at substitutional dimer sites
consistently with the experimental observation that silicon atoms
in Si-DLC are surrounded by carbon atoms, and not by oxygen or other
silicon atoms.\cite{iseki,palshin} 
A density of Si atom corresponding to 8.3 \% is realized, which corresponds to
a typical dopant concentration in Si-DLC.\cite{MGkim,mori2,dohda,hatada}
Figure 1 shows subsequent snapshots of 
the system dynamics 
under normal pressure of 5 GPa and shear stress of 1.25 GPa 
(the entire simulated trajectory is displayed in the movie of
Supplemental data):
after the onset of sliding, one of the substitutional Si atom captures 
a hydroxyl group from a confined water molecule,
and a similar event happens immediately afterwards at another Si-sites.
OH adsorption occurs also at unsaturated C-sites, but 
with lower frequency.  
To clarify the effect of Si doping,
we repeat the simulations maintaining the same
tribological conditions
and interfacial hydrogen coverage, but without the incorporation of any Si atom.
Figure 2 shows the
number of adsorption events 
as a function of time for both the doped and undoped systems.
The comparison of the results clearly indicates that the Si incorporation dramatically increases
the reaction rate for water dissociative adsorption.
\begin{figure*}[htbp]
 \begin{center}
 \includegraphics[width=0.85\linewidth]{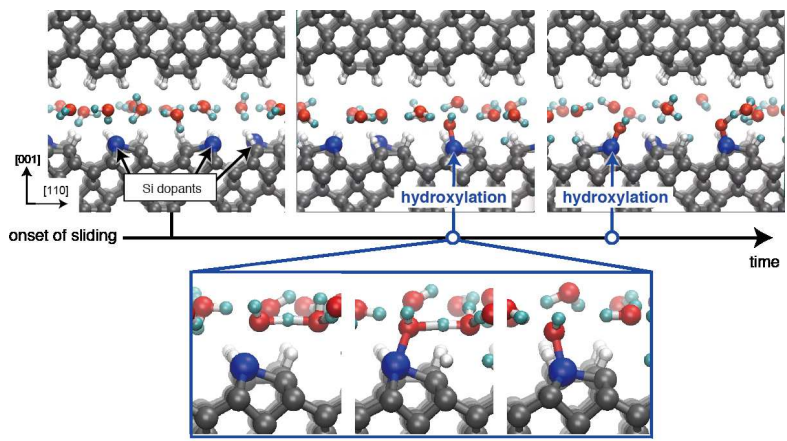}
  \caption{ Snapshots (subsequent in time from left to right) acquired during the molecular dynamics simulation of a partially hydrogenated diamond interface incorporating Si-atoms under 5 GPa pressure and 1.5 GPa shear stress.
  The inset details the process of hydroxyl termination at one Si-site on the surface.
  The red and cyan balls indicate the O and H atoms of water molecules,
respectively.
The white color is used for the H atoms belonging to the initial
  interface termination.
}  \label{pic:snapshots}
 \end{center}
\end{figure*}

\begin{figure}[htbp]
 \begin{center}
 \includegraphics[width=0.60\linewidth]{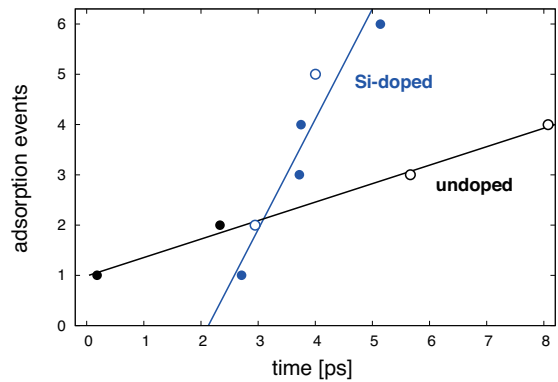}
  \caption{ Number of the adsorption events on the Si-doped
  and undoped surfaces during the dynamic
  simulations at 5 GPa normal pressure and 1.25 GPa shear stress.
Each surface contains six unterminated sites.
The closed and open circles indicate H and OH adsorption,
  resplectively. The solid lines are guides for eyes.
}  \label{pic:reaction_rate}
 \end{center}
\end{figure}

To verify the robustness of the result,
we repeat the sliding simulation for both the systems of Fig.
2,
but with different shear stresses of
1.25 GPa, 0.88 GPa and 0.0 GPa.
The different adsorbate terminations of the surface sites observed during the simulations
are reported in Table 1. 
The result shows
that the Si sites present a
probability to adsorb hydroxyl groups five time higher than
 the C sites, indicating that the Si dopants function
 as catalytic sites for hydroxylation.

\begin{table}
 \begin{center}
\caption{Distribution of the adsorbates attached to 
the initially clean  Si and C surface sites
at the end of six MD simulations 
10 ps long.
The simulatons are carried out under normal pressure of 5 GPa and
shear stresses of 1.25 GPa, 0.88 GPa and 0.0 GPa. Both Si-doped and undoped surfaces are considered. 
}
\label{tab:1}       
\begin{tabular}{ccc}
\hline\noalign{\smallskip}
  & OH termination & H termination \\
\noalign{\smallskip}\hline\noalign{\smallskip}
Si site & 34 \% & 10 \% \\
C site  & 7 \% & 41 \% \\
\noalign{\smallskip}\hline
\end{tabular}
 \end{center}
\end{table}

The acceleration of the hydroxylation reaction by Si dopants observed in the dynamic simulations
 is consistent with 
our previous study
based on static {\it ab initio} calculations, which
shows that
the presence of Si dopants considerably 
decreases the
energy barrier for water dissociation at the surface.\cite{seiji-clelia_static}
Moreover, the analysis of the electronic charge displacements
revealed 
that the larger
polarization of Si-OH bonds with respect to C-OH bonds due
to electronegativity differences, stabilizes hydroxyl adsorption at Si sites,
in agreement with the distribution of the reaction products reported in
 Table 1.
The present dynamics simulations, supported by the analysis of the energetics and thermodynamic
stability of the surface termination,\cite{seiji-clelia_static}
definitely prove that water dissociation 
is promoted by the Si-doping and hydroxylation most likely occurs at Si sites.

\subsection{Formation and function of a water boundary layer}

Having elucidated the mechanisms
of surface hydroxylation in tribologiacal conditions, 
we analyze the effects of this surface termination on the structure and dynamics
of the undissociated water molecules
 that remain confined at the interface. 
To this aim we calculate 
the radial distribution functions (RDF) for the 
O$\cdots$H
bonds 
shown in Fig. 3 (a).
The RDF for bonds between the water molecules and 
hydroxyl groups chemisorbed on the surface
 (green curve) presents two peaks centered around 1.8 \r{A} and 3.3 \r{A}, which correspond to
typical hydrogen-bond distances between
nearest and second nearest neighbor water molecules,
respectively.\cite{waterRDF1,waterRDF2}
 The RDF for surface-molecule bonds is, in fact, very similar 
to that calculated for molecule-molecule bonds within the confined water layer (blue curve).
This similarity indicates that 
the adsorbed OH groups participate to the hydrogen-bond network of the water film.
In other words,
the Si-OH sites anchor water molecules 
by hydrogen-bonding and this link is retained even in severe sliding conditions. 
On the contrary, the RDF profile for the 
H-terminated surface shows the first peak located around 2.5 \r{A}.
This longer bond distance indicates that the hydrogenated surface 
can not establish hydrogen-bonds with the water molecules, which remain 
just weakly physisorbed on it.
The results of this analysis are confirmed by our previous static {\it ab initio}
calculations, which show that the
 physisorption energy of a water molecule is three times lower in the presence of an adsorbed hydroxyl
group than a hydrogen atom.\cite{seiji-clelia_static} 
\begin{figure}[htbp]
 \begin{center}
 \includegraphics[width=0.5\linewidth]{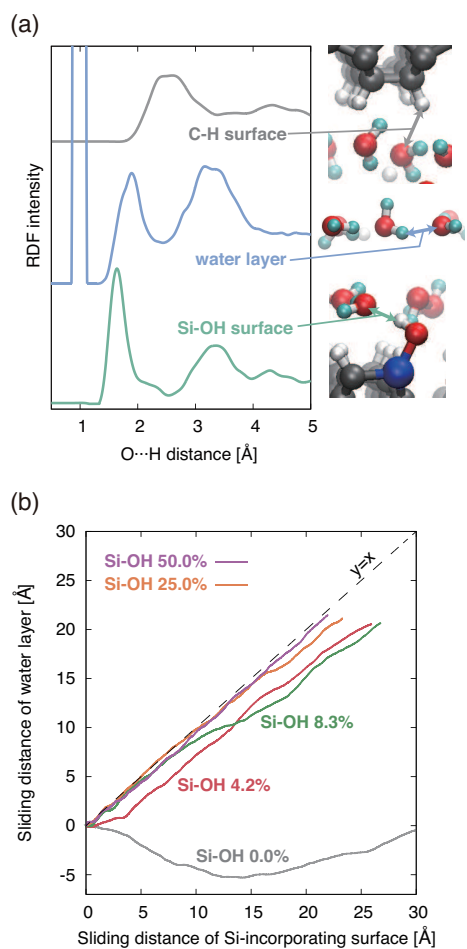}
  \caption{ (a) Radial distribution function of the 
O$\cdots$H bonds
  indicated by the arrows on the left side of the picture, where the ball-and-stick structures
  of the Si-incorporating surface (8.3 \%  Si-OH), water boundary layer, and hydrogenated
  countersurface are represented. 
  (b)
 Sliding distance covered by the water boundary layer as a function of the 
 distance covered by the Si-incorporating surface. Different concentrations of dopants are considered.
} \label{pic:wlpos-rds}
 \end{center}
\end{figure}

The hydrogen bond network established between the hydroxylated surface
and the confined water film affects
the sliding dynamics
of the latter. To clarify this issue, we repeat the sliding simulation
above described by changing the amount of hydroxyl groups adsorbed
on the lower surface, being the upper surface fully hydrogenated and dragged 
with a force of same magnitude and opposite direction.
Figure 3 (b) shows 
the sliding distances covered by the center of mass of the water boundary layer 
as a function of the sliding distance of the lower surface, where the 
density of Si-OH groups ranges from 4.2 \% upto 50 \%.
We can see that the water layer is dragged by the lower surface only if
the latter
is hydroxylated, while in the case of fully hydrogenation (Si-OH 0 \%),
 the confined water film fluctuates around its original position,
without any preferred direction.
The higher the concentration of hydroxyl group at the carbon surface,  
the more tightly the latter binds the
water layer; e.g., the plot for the highest density of OH
groups shows that the water layer slides coherently with the hydroxylated surface.
Nevertheless,
even in the low density case, 4.2 \% OH,
the water molecules are attracted by the slyding surface.

\begin{figure}[htbp]
 \begin{center}
 \includegraphics[width=0.70\linewidth]{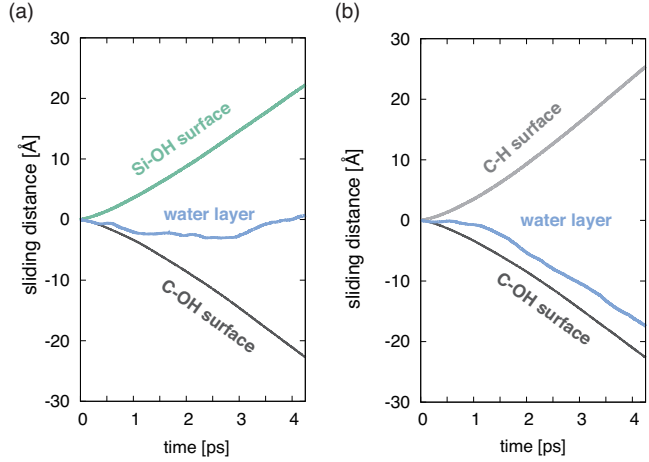}
  \caption{ Sliding distances for water layers confined between two surfaces containing the same amount of hydroxyl groups
  sliding in opposite directions. (a) The same simulation is repeated
  considering a fully hydrogenated countersurface.
(b) The dynamic simulations are performed under 5 GPa normal pressure and
  1.25 GPa shear stress.}\label{pic:wlpos-doping}
 \end{center}
\end{figure}
%

 We conclude by observing that a C-OH terminated-surface has a similar capability of 
water attraction as the Si-OH terminated one. This is elucidated by Fig. 4, 
which shows that doped and undoped surfaces containing the same amount of hydroxyl groups 
drag the water boundary layer with the same intensity. Indeed, the water layer remains
almost at rest during the relative sliding of the two surfaces in opposite directions (panel a).
On the contrary, the water boundary layer follows the C-OH surface if a
 fully hydrogenated counter-surface is considered (panel b), as it happened for the
Si-OH terminated surface of Fig. 3(b). 

\section{Discussion}

One of the proposed explanations for the low friction coefficient of DLC
is that a graphite-like, sp$^2$ carbon tribolayer, 
generated by rubbing the film against a solid countersurface,
decreases the interfacial shear strength.\cite{rabbani,YLiu} 
From this point of view, the mechanism
of silicon doping
in reducing the DLC friction is
regarded as peculiar because it cannot conform to
the above explanation; it is, in fact, well known that the silicon incorporation
decreases the sp$^2$/sp$^3$ ratio of the carbon matrix.\cite{varma,iseki}
Therefore, 
researchers 
introduced another possible explanation for the observed friction reduction 
associating it to the
transfered film of silicon-rich oxide often detected in 
spectroscopic analyses after the sliding tests.
Since the oxides are formed in the presence of water vapor, it 
has been hypothesized that
 water
hydrates the silicon-oxide wear debris and transform them in silica-sol, which may function as
 a sort of liquid-lubricant.\cite{oguri,MGkim,Fzao,scharf}
The silica-sol mechanism was originally proposed to explain friction tests for
SiC in water, during which silicon oxides, continuously generated and removed,
form an extremely smooth, hydrated surface.\cite{woydt,chen2001,chen2002}
However, the production of wear particles, as assumed by this mechanisms,
not always takes place, as for example
in films incorporating a low amount of Si dopants that
provide low friction and wear simultaneously.\cite{dohda,varma}
Sen {\it et al.} performed friction tests to show that
neither Si nor silica glasses
exhibited lower friction than Si-DLC, and pointed out that 
the silicon element must be included within the amorphous
carobn matrix to provide excellent frictional properties.\cite{sen}
\par In parallel to the above observations, 
several authors have reported experimental evidences,
such as derivatization X-ray photoelectron
spectroscopy and nuclear magnetic resonance,\cite{mori2,takahashi,kato,XWu2} that water directly reacts at 
the Si sites incorporated in DLC, producing Si-OH groups. 
Kato {\it et al.}
identified a water layer of thickness ranging from 
1 nm to 4 nm on the Si-DLC surface 
by using  the spectroscopic ellipsometry,
and proposed that
this water layer lubricates the Si-DLC film.\cite{kato}
Recent resonance shear measurements 
revealed that
the water confined between hydroxylated silica surfaces 
forms an ice-like structure
that increases the lubricity of the surface.\cite{kasuya} 
\par The present {\it ab initio} MD simulations support the 
experimental observations pointing at the formation of a water layer 
that may function as a boundary lubricant.
Figure 3 (b)
indicates that even
a small concentration of Si-OH, $\sim$ 4\%, 
is able to attract water molecules and drag them during sliding.
The effectiveness of a small dose of Si was also reported by
experimental works, where the 
the friction reduction obtained by Si 
incorporation was observed up to a dopant concentration of 5 \%.\cite{MGkim,dohda}
The RDF analysis in Fig. 3 (a)
reveals that 
relatively strong hydrogen bonds anchor the thin water layer
to the hydroxylated surface.
Thereby,
it is reasonable to expect that 
the water boundary layer will not be easily squeezed out from the aperity contacts, 
preventing direct solid-on-solid contact and resulting in 
extremly low friction.
On the contrary, the hydrogenated surface 
weakly attracts water molecules, which may be expelled from the interface in 
severe tribological conditions.
This explanation is consistent with and can account for low-friction phenomena
reported in previous studies. \cite{sirghi,IBkim,CMatta,LPottuz,WHabchi,kato,kasuya,washizu}

Our analysis suggests that 
the main role of Si dopants is to increase the
density of hydroxyl groups at the surface
by promoting its hydroxylation. This 
is supported by experiments reporting that
a very small partial pressure of water, $10^{-3}$ Pa, 
is sufficient for the hydroxylation of Si-DLC,\cite{sen}
while undoped DLC loses their water adsorption capability at such low
level of moisture.\cite{konca}

In dry or hydrogen atmosphere,
the H-terminated carbon films 
posses unmeasurably-low friction coefficient of about 0.01 or
lower,\cite{jpcc-diamond,AErdemir}
which originates from the smooth 
potential energy surfaces of
the fully-passivated interface.\cite{righiprb2007,righilangmuir}
Surface passivation leads to extremely low adhesion
and is able to keep the surfaces
apart even under load, thus
preventing the formation of covalent bonds 
and atomistic locking
across the interface.\cite{jpcc-diamond,
zilibotti,righiprb2007,righilangmuir,mhirano0} 
In the case of atomically smooth surfaces, H-passivation has been found
to be more effective than that passivation by water molecules.\cite{jpcc-diamond,AErdemir}
Hydrogen
bonding between the surfaces having the same hydroxyl termination can,
in fact, increase the corrugation of the potential energy surface.\cite{jpcc-diamond}
The mechanism of surface hydroxylation followed by the formation of a
water boundary layer, which has been described in the previous section,
provides a microscopic explanation for the observed increase of
hydrophilic character of Si-doped DLC.\cite{yi,borisenko}
An increase of surface
wettability affects lubricity in many general
situations, where for example the contacting surfaces have a different
hydrophilic character and are not atomically smooth.\cite{borruto}


A very important insight from our analysis
is that the formation of a confined water layer triggered by 
the surface hydroxylation 
can be an universal friction mechanism for the extreme slipperiness of
carbon films. 
Figure 4 indicates, in fact, that
once hydroxylated, 
both undoped and Si-doped carbon surfaces display an
identical capability of attracting water.
This result is in agreement with recent
AFM measurements for undoped DLC films that 
exhibit high lubricity in the presence of adsorbed
water.\cite{sirghi,IBkim}

We conclude the discussion with a comment on experimental observations
that show an
increase of friction in Si-DLC at high levels of humidity.\cite{scharf,hioki}
As stated above, a Si concentration of 5\% has been proven to be
sufficient to achieve low friction and the friction
performance did not improve significantly by adding more Si dopants.
If one performed a severe friction test of 
a densely-doped Si-DLC 
in high humid atmosphere,
the wear would evolve considerably due to the 
silicon-rich oxidation and hydration,
and the eroded surface would increase the friction
coefficient during the test.
Another possibility is related to the viscoelastic property of the
confined water boundary layer.
As pointed out by Sirghi {\it et al.},
the effects of an adsorbed layer of water molecules 
 is not always positive for the friction performance, 
because the water layer confined between surfaces terminated with
highly-concentrated hydroxyl groups
can form a condensed
solid
structure with higher shear stress.\cite{sirghi,IBkim}
Indeed, amorphous solid water is well know to form several structures
with different densities.\cite{JLFinney,TKondo}
In both the cases, the density of the hydroxyl groups is a
key parameter to control the friction properties of the carbon-based films.

\section{Conclusion}
To elucidate the atomistic mechanism underlying the extreme
lubricity of carbon-based films, 
we perform large scale {\it ab initio} molecular dynamics
simulations of 
sliding interfaces of both undoped and Si-doped diamonds interacting with water.
We find that the Si dopants act as catalytic sites for 
surface hydroxylation.
The adsorbed hydroxyl groups participate 
 in the hydrogen bond network
of molecular water confined at the interface, thus anchoring a thin
water film to the surface.
This two-step process well explains the enhancement of
hydrophilicity observed upon doping DLC by silicon. The increased
wettability most likely results in an increased lubricity of DLC
coatings.
On the basis of the observed sliding dynamics, 
we, indeed, expect that this water film
is not easily squeezed out from contacting surfaces, preventing
their solid-on-solid contact and making friction and wear very low. 
Interestingly, we show that
pure carbon-based films once hydroxylated present 
the same capability of binding
the nanoconfined water layer as Si-doped surfaces. 
The density of the hydroxyl groups is a
key parameter to control the frictional properties 
in different environments. 
Since this parameter can be handled 
by suitable surface treatments\cite{kasuya} and doping techniques,\cite{sanchez,donnet,grill}
the results described 
in the present paper
will provide important guidelines for 
carbon-material design in
several fields related to energy-saving applications.

\section*{Acknowledgments}
We thank the CINECA consortium for the availability of high
performance computing resources and support through the ISCRA-B TRIBOGMD
project.

\section*{Supplementary data}
Supplementary data associated with this article can be found, in the
online version, at http://

%

\end{document}